\documentclass[12pt]{elsart}
\usepackage{graphicx}
\usepackage{amsmath}
\usepackage{amsfonts}
\usepackage{amssymb}
\usepackage{bbm}
\usepackage{tabularx}
\usepackage{epsf}

\begin{document}

\runauthor{Miceli and Susinno}
\begin{frontmatter}
\title{Ultrametricity in Fund of Funds Diversification}
\author[sapienza]{M. A. Miceli}
\author[fp]{G. Susinno}

\address[sapienza]{Department of Public Economics University of Rome ``La Sapienza''}
\address[fp]{Finance-and-Physics Independent Advisor in Computational and
  Quantitative Finance. E-mail contact: susinno@finance-and-physics.org}

\begin{abstract}
\par Minimum market transparency requirements impose Hedge Fund (HF) managers to use the
statement declared strategy in practice. However each declared
strategy may actually origin a multiplicity of implemented
management decisions. Is then the "actual "strategy the same as
the "announced" strategy? Can the actual strategy be monitored or
compared to the actual strategy of HF belonging to the same
"announced" class? Can the announced or actual strategy be used as
a quantitative argument in the fund of funds policy? With the
appropriate metric, it is possible to draw a minimum spanning tree
(MST) to emphasize the similarity structure that could be hidden
in raw correlation matrix of HF returns.
\end{abstract}
\end{frontmatter}

\section{Introduction}

\par Before confined to off shore destinations, Hedge Funds (HF) are
managed portfolios under the least number of constraints. HF,
enjoying more freedom than traditional institutional funds, could
historically achieve better returns under the same market
conditions: for example, short selling means the possibility to
make profit out of a bear market. Investing in a HF is a trust
contract between the investor and the fund manager's strategy.
However complete freedom in managing investors money is hard to
sell. In order to require a minimum transparency between the HF
manager strategy and investors, national regulatory institutions
ask the hedge funds to declare the adopted policy in the
statement. Practice has shrank the ''types'' of style management
in some, by now, well known classes. Why do we care about hedge
fund type of style management? Because a growing market practice
tends to classify HF as normal assets, hence as objects of a
further investor portfolio optimization. Unawareness of HF
peculiarities and biases may hide the risk of being fooled by
randomness.  For a nice introduction to HF consult
e.g.~\cite{Lhabitant}.
\par Data on HF performances present more shortcomings than
asset time series: HF Net Asset Value (NAV) are quoted just once a
month. Give the highly proprietary character of applied
strategies, which are jealously guarded, the only public data are
about indexes grouping HF by style. These indices contain a set of
disappearing and appearing HF, inducing an unavoidable "survivor
bias".
\par In this first paper we focus our attention on the
characterization of the applied strategy, in particular our data
set shows that strategies plays in HF a role similar to market
sectors for traditional assets. Indeed we apply combinatorial
optimizing techniques to analyze those characteristics that
cluster funds together and those that diversify them. Finally we
show how this approach can be used to monitor and select HF
investments highlighting those individual HF with a tendency to
depart from their stated investment strategy.
\section{The Data}
\par We use $N$ single HF NAV time series of $T$
synchronous observations, where:
\begin{itemize}
\item  Time: Jan-1999 - Jan-2003 of 49 monthly observations;
$T=49$
 \item  Assets:
\begin{itemize}
\item 62 Funds with strategies reported in Table
(see~\cite{Lhabitant} for strategies definition ).
\item 5 Market
Indices: MIB30, DJ, HSI, NDQ, FTSE, SP500
\end{itemize}
\end{itemize}
\section{Noise and Signal in Correlation}
\par   Rank reduction for correlation matrices
is usually obtained via a standard zeroed-eigenvalues reduced-rank
approximation~\cite{Brigo,Rebonato2}. As in~\cite{Stanley1} we
identify the significant eigenvalues of correlation matrix by
selecting those which depart from the spectrum of a same size
symmetric random matrix. Random Matrix Theory (RMT) offers a way
to clean the matrix from the random components~\cite{metha}. The
``cleaned'' correlation matrix is then used to determine Euclidean
distances between funds.
\par We first normalize monthly returns for the all series,
then compute the equal time cross-correlation $N~\times~N$ matrix
$C$. Problems in measurement (see e.g.~\cite{Stanley1}) are: (i)
non stationarity of the matrix as market conditions change; (ii)
the finite length of time series available to estimate cross
correlations introduces ''measurement noise''. As $T$ increases to
avoid problem (ii), problem (i) increases. From both sources (i)
and (ii) we get random contributions into the correlation matrix.
\par
\begin{figure}[h]
\begin{center}
$\begin{array}{c@{\hspace{1in}}c} \multicolumn{1}{l}{\mbox{\bf
(a)}} &
    \multicolumn{1}{l}{\mbox{\bf (b)}} \\ [-0.53cm]
\epsfxsize=3in \epsffile{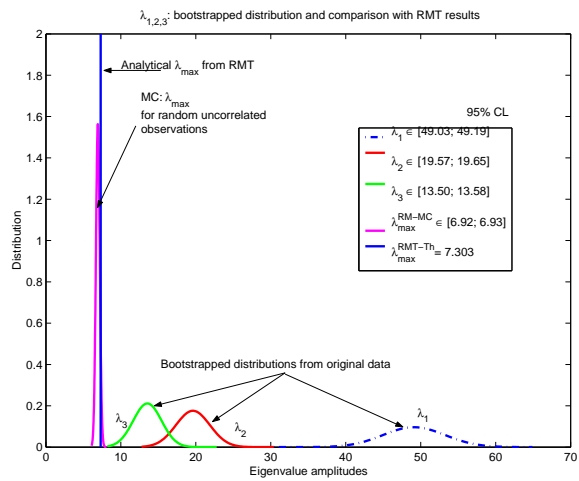} &
    \epsfxsize=3in
    \epsffile{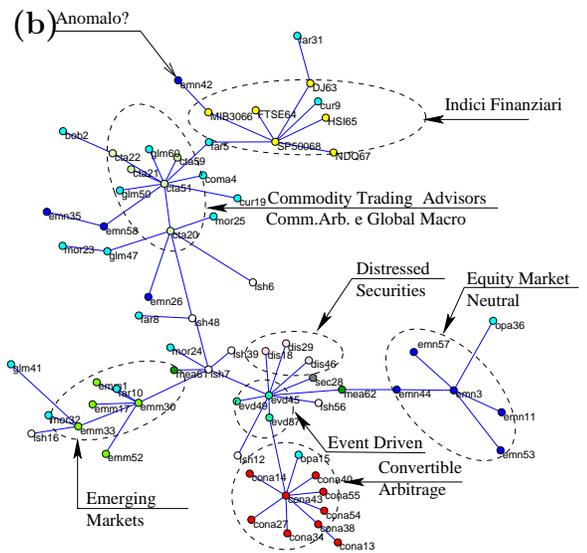} \\ [0.4cm]
\end{array}$
\end{center}
\caption{{\bf (a)} Highest eigenvalues distribution compared to
$\lambda_+$ given by RMT results for uncorrelated returns; {\bf
(b)} Minimum Spanning Tree for HF Strategies.}
\label{figtest-fig}
\end{figure}

 We test the eigenvalues of the matrix C against the null hypothesis given by
eigenvalues generated by a same size symmetric random matrix. In
the limit $N,T\longrightarrow\infty$ with a fixed radio $Q=T/N\geq
1$, the spectral density is given by~\cite{Sengupta}:
\begin{equation}
 P_{rm}\left(  \lambda\right)
=\frac{T}{2\pi}\frac{\sqrt{\left( \lambda
_{+}-\lambda\right)  \left(  \lambda-\lambda_{-}\right)  }}{\lambda}%
,\qquad\text{where }\lambda_{\pm}=1+\frac{1}{Q}\pm2\sqrt{\frac{1}{Q}}%
\label{eq:spectrum}
\end{equation}
The three highest eigenvalues of the correlation matrix spectrum
for 62 HF and 5 market indices, is compared to the highest
eigenvalue $\lambda_+$ given by eq:~\ref{eq:spectrum}. To gain
hindsight on the measurement accuracy for the three highest
observed eigenvalues $\lambda_1,\lambda_2,\lambda_3$ we have
determined the bootstrapped distribution of their estimators.
Since eq:~\ref{eq:spectrum} is valid for
$N,T\longrightarrow\infty$ s.t. $Q=T/N\geq 1$ is fixed, we test
the finite size effect on $\lambda_{max}=\lambda_+$ determination.
Results are reported on Fig:~\ref{figtest-fig}(a), showing that
the three highes eigenvalues of the observed correlation matrix
are not compatible with the Random Matrix Hypothesis. In the
following we will use the rank-reduced matrix $\overline{C}$
obtained by zeroing all the eigenvalues lower than $\lambda_3$.
Since zeroing eigenvalues has altered the diagonal we may consider
$\overline{C}$ as a covariance matrix, the associated correlation
matrix can be recovered as follow:
\[
{\hat{C}}_{ij}=\frac{\overline{C}_{ij}}{\sqrt{\overline{C}_{ii}\overline
{C}_{jj}}}%
\]
\section{Minimum Spanning Tree: How to Visualize Dependencies}
\par We consider HF time series as points in $\mathbbm{R}^{T}$ Euclidean
space. The method is based on a similarity concept expressed by
distance: similarity grows as distance shrinks. A frequently used
measure is the Euclidean distance $d_{ij}$ between two normalized
time series, that can be calculated starting from the linear
correlation coefficient~\cite{Mantegna,Rammal}.
\begin{equation}
d_{ij}^{2}=\left\|
\overset{\thicksim}{\mathbf{S}_{i}}-\overset{\thicksim
}{\mathbf{S}_{j}}\right\|  ^{2}=\sum_{k=1}^{n}\left(
S_{ik}-S_{jk}\right)^{2}=2\left(1-\hat{C}_{ij}\right)
\label{eq:dist}
\end{equation}
\par Given distances between all funds and by considering each fund
as a node of a tree structure, it is possible to determine a data
representation known as Minimum Spanning Tree (MST). It is the
unique graph that connects all nodes with the minimum extension.
The minimum spanning tree is obtainable through simple algorithms
[PA01]. Distance minimization between tree nodes allows for a
natural HF classification in clusters containing elements with
similar realizations.
\par In our example, using return time
series of 62 hedge funds in the period January 2000 to January
2003, we determined the associated MST
(Fig.~\ref{figtest-fig}(b)). We note how the clustering process
leads to an economic meaningful classification of strategies and
how anomalies can be promptly detected. For some clusters (cona,
gta/glm, emn, emm) there exist a fund that acts as a reference
center (cona 43, cta51, emm30, ), while strategies as "Long-Short"
do not show peculiar characteristics, coherently with
heterogeneity and discretion in this management style. This
observation may suggest that the less discretionary management
policies are (for examples, those following mathematical models
implemented by software decisions), the more similar corresponding
returns are.
\section{Conclusions}
\par We observe a broad coherence between the usual qualitative classification of HF
and the phenomenological classification deduced from historical
data of HF returns. Albeit managers discretionality, macro
strategies seems to share enough common points in the realized
returns to be quantitatively classified. In a universe where
transparency towards investors is low, where operative strategies
are protected, data mining and classification instruments allow to
extract maximum information from available data and to verify
whether the declared strategies in the fund statement are
verified. In particular, anomalous nodes with respect to the
reference cluster, let identify funds that require a deeper
investigation. Beyond qualitative control, these results are
useful to maximize portfolio diversification, by selecting funds
belonging to different clusters and by paying special attention to
those characterizing central nodes. Moreover, those
characteristics that identify a cluster help to define the actual
benchmark for a given strategy. Finally, the tree structure offers
an objective basis to extract economic conclusions, portfolio
selection and control. By extracting the structure hidden in large
correlation matrices, trees are easier to interpret than
inspection of large correlation matrices!

\bibliographystyle{alpha}
\bibliography{hf-div-gab}
\end{document}